\documentclass[aps,prl,twocolumn,superscriptaddress,showpacs,showkeys]{revtex4}
\usepackage{epsfig}
\usepackage{graphicx}
\usepackage{bm}
\usepackage{natbib}
\usepackage{dcolumn}
\usepackage{amsmath}
\usepackage{rotating}
\usepackage{multirow}

\begin{document}

\title{Complex networks renormalization: flows and fixed points}

\author{Filippo Radicchi}
\affiliation{Complex
  Systems Lagrange Laboratory (CNLL), ISI Foundation, Torino, Italy}
\author{Jos\'e Javier Ramasco}
\affiliation{Complex
  Systems Lagrange Laboratory (CNLL), ISI Foundation, Torino, Italy}
\author{Alain Barrat}
\affiliation{Complex
  Systems Lagrange Laboratory (CNLL), ISI Foundation, Torino, Italy}
\affiliation{Laboratoire de Physique Th\'eorique (CNRS UMR 8627),
Universit\'e de Paris-Sud, France}
\author{Santo Fortunato}
\affiliation{Complex
  Systems Lagrange Laboratory (CNLL), ISI Foundation, Torino, Italy}

\begin{abstract}

Recently, it has been claimed that some complex networks are self-similar under a convenient renormalization procedure.
We present a general method to study renormalization flows in graphs. We find that the
behavior of some variables under renormalization, such as the maximum number of connections of a
node, obeys simple scaling laws, characterized by critical exponents. This is true for any class
of graphs, from random to scale-free networks, from lattices to hierarchical graphs. Therefore, renormalization
flows for graphs are similar as in the renormalization of spin systems. An analysis
of classic renormalization for percolation and the Ising model on the lattice confirms this analogy. Critical
exponents and scaling functions can be used to classify graphs in universality classes, and to uncover
similarities between graphs that
are inaccessible to a standard analysis.

\end{abstract}
\maketitle

Generally speaking, an object is self-similar if any part of it,
however small, maintains the general properties of the whole
object. Self-similarity is a characteristic feature of
fractals~\cite{mandelbrot} and it expresses the invariance of a
geometrical set under a length-scale transformation.  Many complex systems
such as the World-Wide-Web (WWW), the Internet, social and biological
systems, have a natural representation in terms of graphs, which often
display heterogeneous distributions of the number of links per node
(the degree $k$)~\cite{Barabasi02,Dorogovtsev03,newman04,pastor04,boccaletti06}. 
These distributions can be described by a
power law decay, i.e.  are scale-free: they remain invariant under a
rescaling of the degree variable, suggesting that suitable
transformations of the networks' structure may leave their statistical
properties invariant. Since graphs however are not embedded in
Euclidean space, a standard length-scale transformation cannot be
performed. The concept of length can only be defined in the
graph-theoretical sense of the number of links along any shortest path
between two nodes. In this context, Song et al.~\cite{song05} proposed
to transform a network by means of a box-covering technique, in which
a box includes nodes such that the distance between each pair of nodes
within a box is smaller than a threshold $\ell_B$.  After tiling the
network, the nodes of each box and their mutual links are replaced by
a supernode: supernodes are connected if in the original network there
is at least one link between nodes of their corresponding boxes. This
defines a renormalization transformation $R_{\ell_B}$. For some
real networks, such as the WWW, social, metabolic and
protein-protein interaction networks, a few iterated applications of
this procedure seem to leave their degree distribution invariant,
which led to the claim that they are self-similar~\cite{song05}.
Other networks, such as the Internet, are instead found not to be
self-similar under $R_{\ell_B}$.

Iterated applications of $R_{\ell_B}$ generate renormalization flows
in the space of all possible graphs. Studying the behavior of such
flows is crucial: the existence of possible fixed points of the
transformation would allow to identify universality classes of
networks, much like it happens for second-order phase transitions in
statistical physics~\cite{stanley}. This could offer a natural way to
classify graphs and uncover unknown similarities. In this paper we
perform a systematic study of the renormalization transformation
$R_{\ell_B}$, its flows and fixed points.

We denote a generic graph of $N_0$ nodes and $E_0$ links by $G_0$ and the
renormalization transformation by $R$, for simplicity. A series of $t$
successive transformations $R$ on $G_0$ leads to the graph $G_t=R^t(G_0)$,
with $N_t$ nodes and $E_t$ links.  Finite size effects are strong especially
in heterogeneous networks, where boxes built around large degree nodes (hubs)
determine a considerable contraction of the system at each step. Such effects
may perturb the analysis of the renormalization flow, which therefore has not
been investigated so far. We have devised a general procedure that overcomes
this difficulty and allows to study the renormalization flows.

Tiling a network means covering it with the minimum number of boxes. We
adopted two popular techniques for box covering: the greedy coloring
algorithm~\cite{song07} (GCA) and random burning~\cite{goh06} (RB).  GCA is a
greedy technique inspired by the mapping of the problem of tiling a network to
node-coloring, a well known problem in graph theory~\cite{bollobas}.  In RB,
boxes are spheres of radius $r_B$ centered at some seed nodes, so that the
maximal distance between any two nodes within a box does not exceed $2r_B$. 
The correspondence between the two methods is obtained for $\ell_B=2r_B+1$. The main results 
of our analysis appear robust with respect to the particular adopted box covering technique.

An important characteristic of a network is its largest
degree. We therefore focus on
the variable $\kappa_t=K_t/(N_t-1)$, where $K_t$ is the largest degree of graph
$G_t$. As the number of
renormalization steps $t$ increases, we study the flows of $\kappa_t$
as a function of the relative network size $x_t=N_t/N_0$ ($N_0$ is
the initial network size).  We also study the fluctuations of the variable
$\kappa_t$ along the flow, expressed by the susceptibility $\chi_t=N_0\left( \langle
{\kappa_t}^2\rangle-\langle \kappa_t\rangle^2\right)$; here the averages, denoted by
$\langle \cdot \rangle$ are taken over various realizations of the covering
algorithm.

\begin{figure}[htb]
\includegraphics[width=7cm,height=5cm]{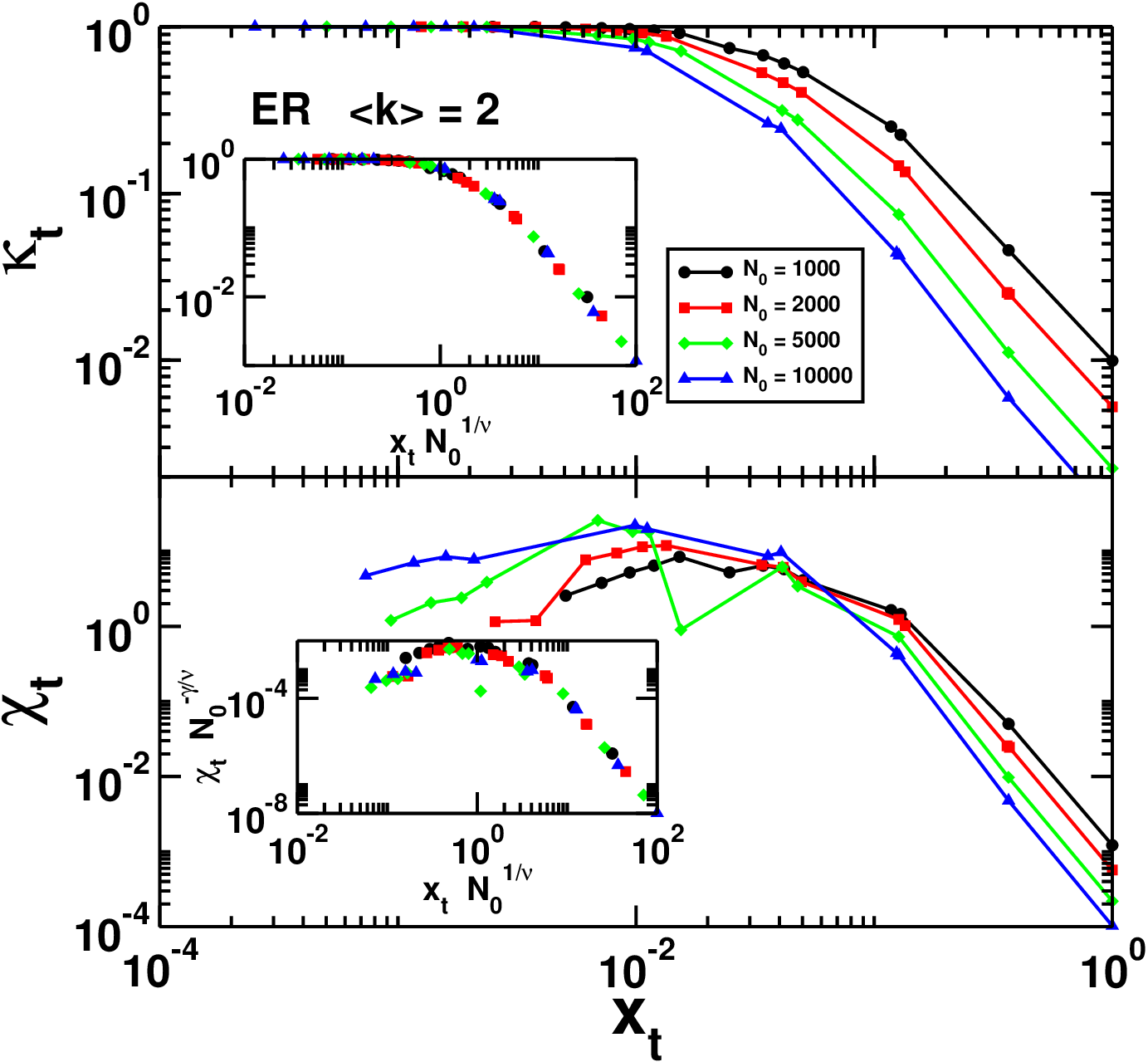}
\caption{\label{fig1} Study of renormalization flows on non-self-similar artificial
  graphs. The box covering was performed with GCA for $\ell_B=3$.
  After $t$ iterations of the renormalization procedure, the
  graph $G_t$ has $N_t$ nodes and $E_t$ links and its maximal degree
  is $K_t$.  The graph is an ER network with average degree $\langle k \rangle=2$. 
  The figures display $\kappa_t=K_t/(N_t-1)$ (top) 
  and $\chi_t=N_0\left( \langle {\kappa_t}^2\rangle-\langle \kappa_t\rangle^2\right)$
  (bottom) as a function of the relative network size $x_t=N_t/N_0$. 
  The insets display the scaling function of the variable $x_tN_0^{1/\nu}$ for $\kappa_t$ and $\chi_t$. 
  Here $\nu=2.0(1)$ and the susceptibility exponent $\gamma=\nu$ (within errors).}
\end{figure}

In Fig.~\ref{fig1}
we see how the variables evolve for an Erd\"os-R\'enyi~\cite{erdos} (ER)
graph with average degree $\langle k \rangle =2$, which thus contains
a giant component and has loops. Such network is not self-similar according to 
box-covering renormalization~\cite{song05}.
The box covering was carried out with GCA, with $\ell_B=3$. 
We find that the functions $\kappa_t(x_t)$ and $\chi_t(x_t)$ are scaling functions
of the variable $x_tN_0^{1/\nu}$, as indicated by the
remarkable data collapse of the insets. 
The scaling relations hold on a very
general ground, namely for 
all the box covering procedures investigated, 
with exponents identifying a narrow set of 
universality classes. In the case of non-self-similar objects the estimates for the exponent $\nu$ are consistent
with the value $2$.
The scaling of the susceptibility curves 
requires another exponent $\gamma$, which 
controls the divergence of the peaks (see inset of Fig.\ref{fig1}, bottom). 
We obtain $\gamma=\nu$ for all graphs and transformations.
\begin{figure}[htb]
\includegraphics[width=7cm,height=5cm]{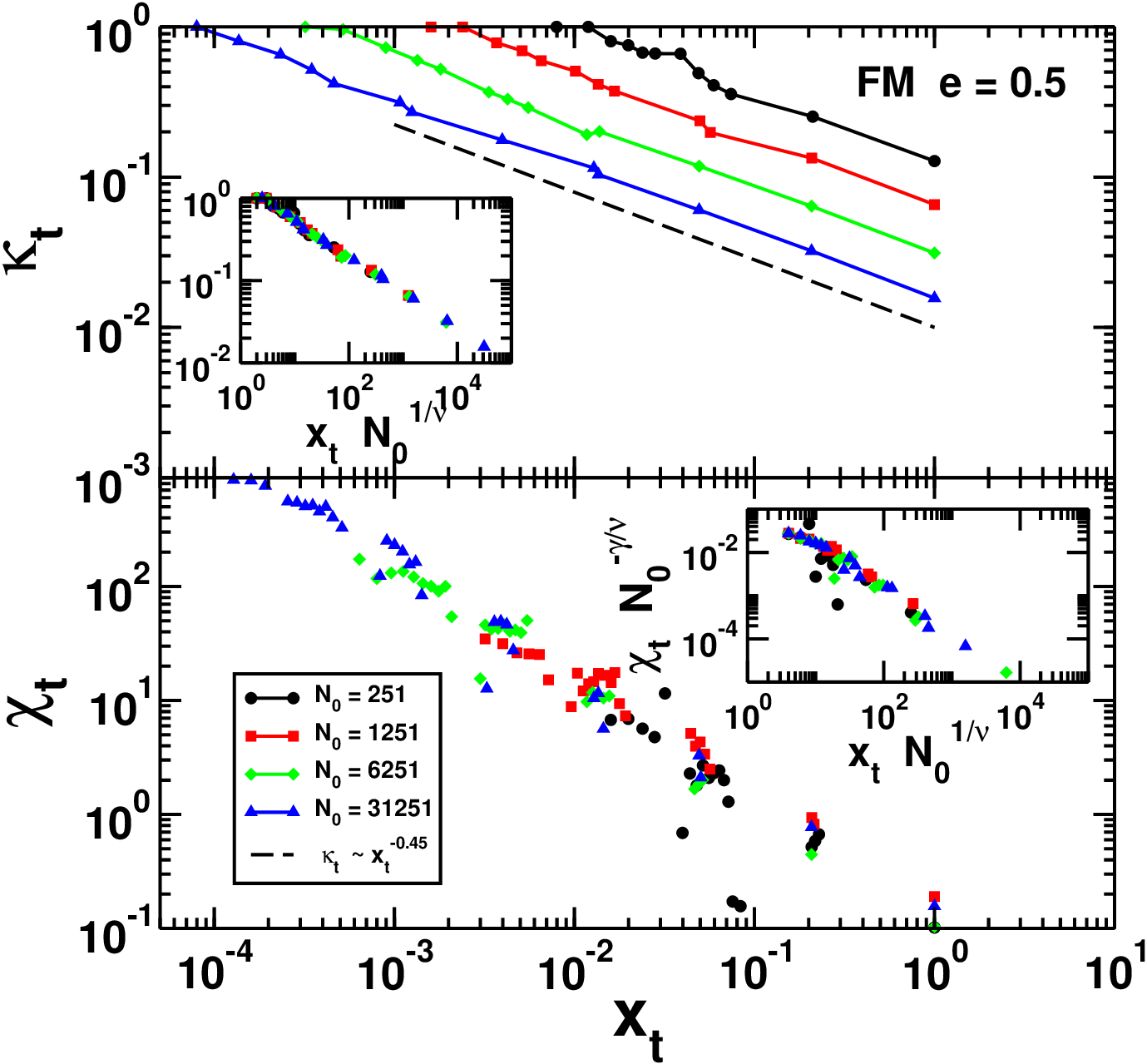}
\caption{\label{fig3} Study of renormalization flows on self-similar artificial
  graphs. The box covering was performed
  with GCA for $\ell_B=3$. The graph is an FM network 
  with $e=0.5$, where $e$ is the probability for hub-hub
  attraction~\cite{song06}.
  The figure displays $\kappa_t=K_t/(N_t-1)$ (top), 
  and $\chi_t=N_0\left( \langle {\kappa_t}^2\rangle-\langle \kappa_t\rangle^2\right)$ 
  (bottom) as a function of the relative network size $x_t=N_t/N_0$. 
  The scaling function 
  of the variable $x_tN_0^{1/\nu}$ for $\kappa_t$ and $\chi_t$ is displayed in the insets.
  The exponent is $\nu=1.05(5)$. The dashed lines 
  indicate the predicted behavior of the scaling function. 
  The scaling function decays with an exponent $-(\beta-2)/(\beta-1)=-0.45$.
  We still find $\gamma=\nu$ (within errors).}
\end{figure}

In Fig.~\ref{fig3} we study the flows for a class of graphs which are self-similar under 
box-covering renormalization: the fractal model (FM)
introduced by Song et al.~\cite{song06}.
The Fractal Model is self-similar by design, as
it is obtained by inverting the renormalization
procedure. At each step, a node turns into a star, with a 
central hub and several nodes with degree one.
Nodes of different stars can then be connected in two ways: 
with probability $e$ one connects the hubs
with each other (Mode I), with probability $1-e$ a non-hub of 
a star is connected to a non-hub of the other
(Mode II). The resulting network is a tree with power law 
degree distribution, the exponent of which depends on
the probability $e$.

This type of graphs maintain their 
statistical features under renormalization. Nevertheless, the scaling behavior is the same we have
observed for non-self-similar graphs, but with different exponents. 
In the case of the FM network it is possible to derive the scaling exponent $\nu$, by
inverting the construction procedure of the graph. In this way one recovers graphs with identical structure at each
renormalization step and one can predict how $\kappa_t$, for instance, varies as the flow progresses.
Since we are interested in
renormalizing the graph, our process is the time-reverse of the growth
described in~\cite{song06}, and is 
characterized by the following relations 
\begin{equation}
N_{t-1}=n\,N_t, \quad
k_{t-1}=s\,k_t, \quad
\beta=1+\frac{\log\,n}{\log\,s},
\label{eq1}
\end{equation}
where $n$ and $s$ are time-independent constants determining the value of the
degree distribution exponent $\beta$ of the network. Here $N_t$ and $k_t$ are
the number of nodes and a characteristic degree of the network at step $t$ of
the renormalization; we choose the maximum degree $K_t$. The initial network has size $N_0$ and shrinks
due to box-covering transformations. In this case, for the variable $\kappa_t$
one obtains
\begin{equation}
\begin{split}
\kappa_t \sim \frac{K_t}{N_t}= \frac{K_0}{N_0}\left( \frac{s}{n} \right)^{-t}=
\frac{K_0}{N_0}\left( \frac{N_t}{N_0}\right)^{-\frac{\beta-2}{\beta-1}}\\
=\frac{K_0}{N_0}x_t^{-\frac{\beta-2}{\beta-1}}\sim (N_0\,x_t)^{-\frac{\beta-2}{\beta-1}},
\end{split}
\label{eq2}
\end{equation}
where we used $s=n^{1/(\beta-1)}$, $N_t/N_0=n^{-t}$ and $K_0 \sim
N_0^{1/(\beta-1)}$, derived from
Eqs.~\ref{eq1}. We see that the scaling exponent $\nu=1$ is obtained
for any value of
the exponent $\beta$. From Eq.~\ref{eq2} we actually get the full shape of the
scaling function, that is a power law: our numerical calculations confirm this
prediction.  We remark that this holds only because one has used precisely the
type of transformation that inverts the growth process of the fractal network.
This amounts to applying the GCA with $\ell_B=3$.

\begin{figure}
\includegraphics[width=7cm,height=5cm]{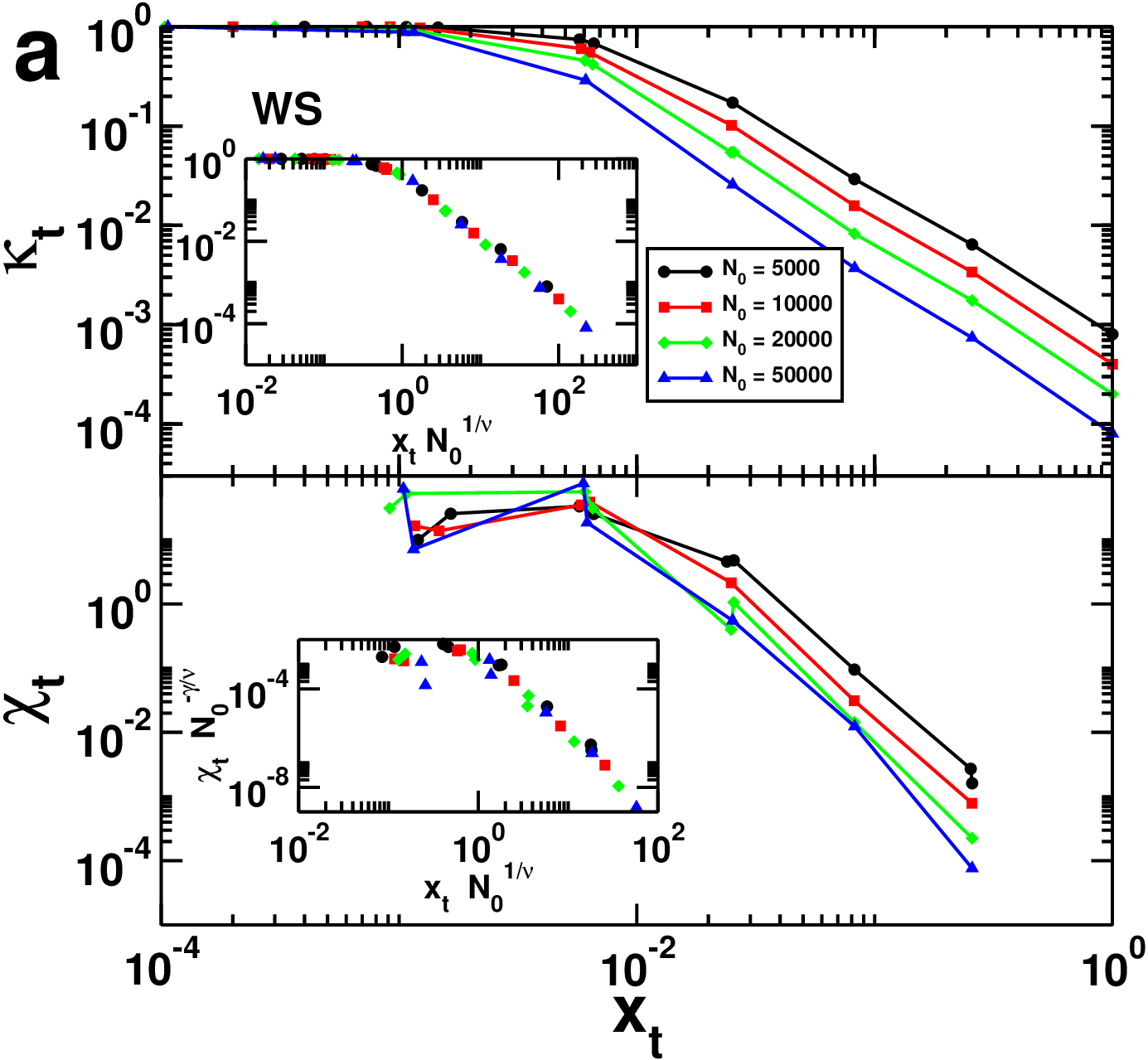}
\includegraphics[width=7cm,height=5cm]{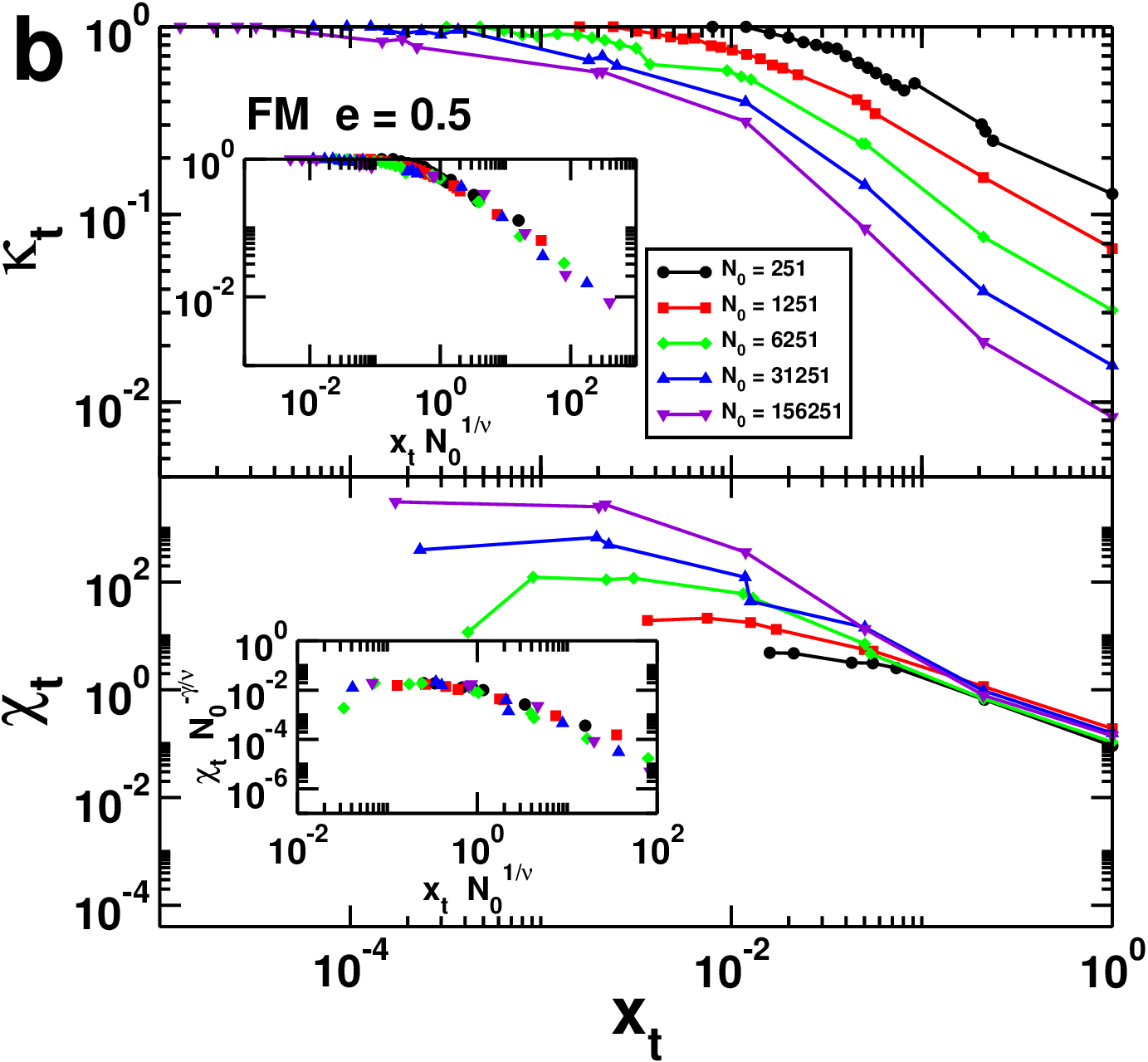}
\caption{\label{fig4} Effect of a small random perturbation on renormalization flows. 
The box covering was performed with GCA, with $\ell_B=3$.
a) WS network with $\langle k \rangle =4$ and a fraction $p=0.01$ of randomly rewired links. 
b) FM network with $e=0.5$ and a fraction $p=0.05$ of added links. 
The figures display $\kappa_t=K_t/(N_t-1)$ (a, b, top), 
and $\chi_t=N_0\left( \langle {\kappa_t}^2\rangle-\langle \kappa_t\rangle^2\right)$
(a, b, bottom) as a function of the relative network size $x_t=N_t/N_0$. We see that 
the exponents are now very different from the unperturbed case: we recover $\nu=2.0(1)$, as in the case of 
non-self similar graphs. The relation $\gamma=\nu$ is still satisfied within errors.}
\end{figure}

Self-similar objects correspond by definition to fixed points of the
transformation. To study the nature of these fixed points, we have repeated
the analysis of the renormalization flows for the self-similar networks
considered, but {\em perturbed} by a small amount of randomness, through the
addition or rewiring of a small fraction $p$ of links.  The
results are shown in Fig.~\ref{fig4} for 
Watts-Strogatz (WS) small-world networks~\cite{watts}
and FM networks. In both cases we
recover the behavior observed for non-self-similar graphs, with scaling
exponents $\nu=2.0(1)$ and $2.0(1)$, which implies that the original fixed points are unstable
with respect to disorder in the connections.
To complete our analysis, 
we have studied the renormalization flows for many other artificial networks,
either self-similar or not,
such as scale-free networks \'a la Barab\'asi-Albert~\cite{BA} or generated with linear preferential
attachment~\cite{DMS}, ER graphs at the threshold for the formation of the
giant component ($\langle k \rangle =1$), hierarchical and Apollonian
networks~\cite{ravasz,apollonian}.  In all cases we have found the same scaling
behavior for $\kappa_t$ and $\chi_t$. We warn that the values of the exponents may 
{\em a priori} also be affected by the specific transformation
adopted, as it happens in real space renormalization for 
lattice models~\cite{stanley}. Still, we find a coherent picture: non-self-similar graphs 
are characterized by exponents consistent with $\nu=2$; self-similar graphs yield different values
for $\nu$.
\begin{figure}
\includegraphics[width=\columnwidth]{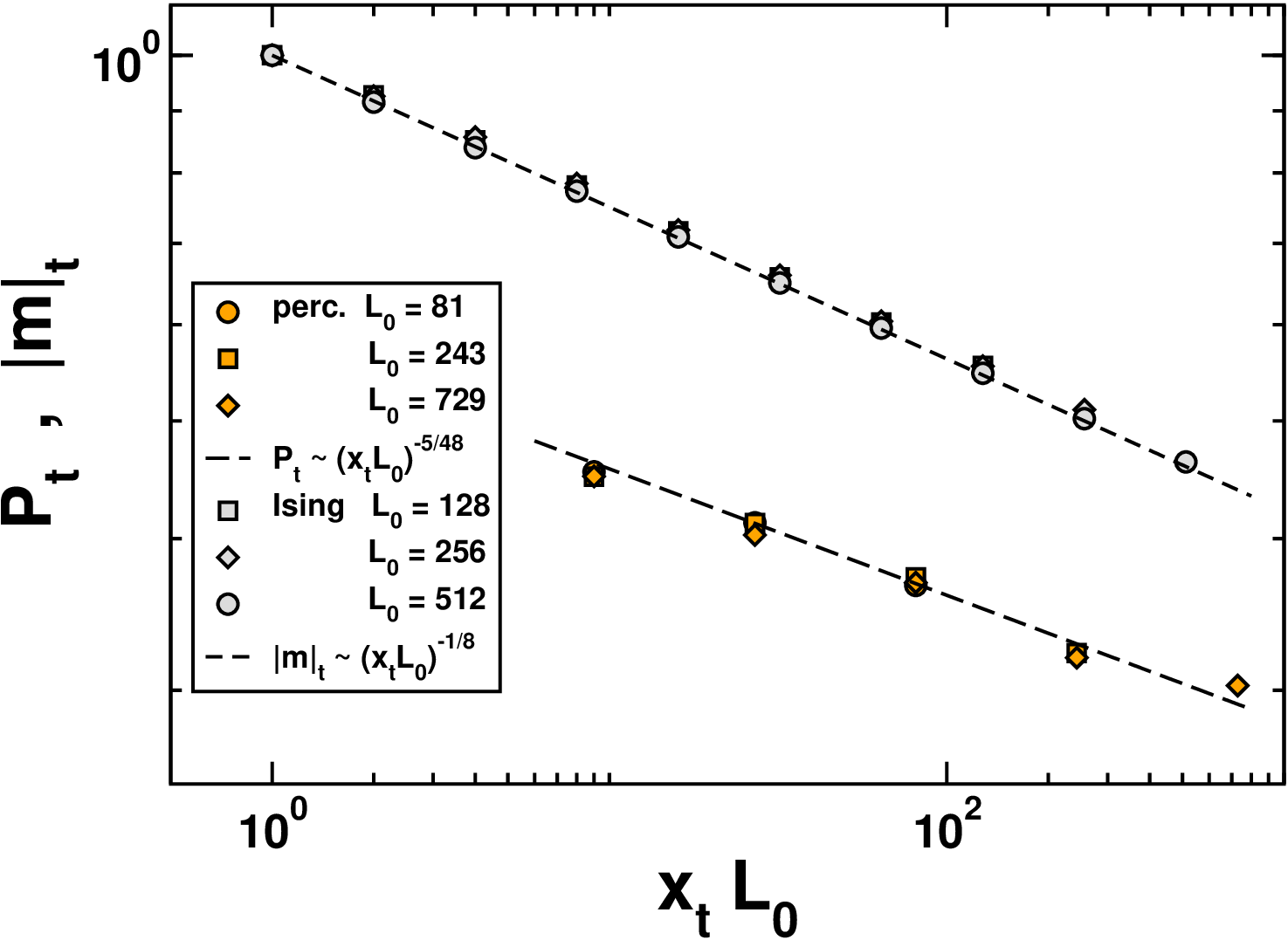}
\caption{\label{fig5} Analogues of our scaling plots for real space renormalization in percolation and the Ising model in two dimensions.
For percolation we use a triangular lattice, and the renormalization reduces the volume by $1/9$ at each step. For Ising we use a square lattice,
with a volume contraction factor of $1/4$. The relative system size $x_t$ now refers to the linear dimension $L$ of the lattices. So,
the values of $x_t$ are multiples of $1/3$ for percolation, of $1/2$ for Ising.
The plot illustrates the flows obtained starting from the critical value
of the control parameter, corresponding to the occupation density $p=0.5$ for percolation
and the temperature $kT=2.269$ for Ising. The two order parameters, the percolation strength 
(relative size of the percolating cluster) and the magnetization, scale with $x_t$. We recover the well known exponents of percolation 
and Ising ($\beta_p/\nu_p=5/48$, $\beta_I/\nu_I=1/8$).}
\end{figure}

The scaling relations we have found are somewhat unusual, as the scaling variable entails 
the relative system volume $x_t$ and not a control parameter.
To disclose the meaning of our scaling, we repeated our analysis for two traditional systems of statistical physics:
percolation and the Ising model in two dimensions. We have applied real space renormalization to percolation configurations
on a triangular lattice, and to Ising configurations on a square lattice. For percolation, a triangular cell is replaced by 
a supernode, which is occupied if the majority of sites of the cell are occupied, empty otherwise
(the procedure is described in~\cite{staufferbook}). For Ising we have applied
a classical majority rule scheme to square cells with four spins. In Fig.~\ref{fig5} we show the relation between the 
order parameter and $x_t$ (which here indicates the contraction in the linear dimension $L_0$), for different initial lattice
sizes, starting from configurations at the critical point. We observe a clean scaling, just as in network renormalization.

The plots are analogues of the standard finite size scaling plots. The order parameter 
scales as $L^{-\beta/\nu}$ ($L$ being the linear dimension of the lattice) at the critical density, which in our case reads $(x_tL_0)^{-\beta/\nu}$
and matches the trend observed in the figure. At variance with finite size scaling, where one always considers configurations of the same system,
here the renormalization may bring the
system to configurations corresponding to the critical state of other
systems in the same universality class, but the scaling holds. We have also 
repeated the analysis starting from system configurations in the subcritical and supercritical phases, in which cases no scaling is observed.

The scaling of Fig.~\ref{fig5} does not give new insight about percolation and Ising, 
as it just reproduces well known exponents. Standard finite size scaling
does the same job, but there one has a control parameter (occupation density, temperature) that allows
to identify the state of the system. In the case of networks, the state is represented by the topology of the system and there is 
no obvious control parameter, so our approach seems the only possibility to extract information about possible critical properties.
The scaling for self-similar graphs in Fig.~\ref{fig3} corresponds to the critical scaling of Fig.~\ref{fig5}.

In conclusion, our results show that
renormalization flows in graphs, as defined by the
box-covering method, display a clear scaling
behavior, opening a new promising research avenue in the field of complex
networks, with close contacts to real space renormalization in lattice models.
Our analysis uses the well-established finite-size scaling and real space
renormalization techniques and could be easily generalized to other possible
renormalization schemes. For a full classification of networks in
universality classes it seems necessary to explore further the robustness 
of the critical exponents under renormalization, and to study the flow
of other variables, which may deliver new interesting scaling functions and
exponents. The analogies we have found with the classic renormalization of 
percolation and the Ising model on the lattice are intriguing and give more insight to our picture. 
Finally, an interesting open question concerns the possibility to
assign real-world networks to specific universality classes. 
This is a challenging issue, as for real graphs a finite-size scaling analysis is not
available because of the uniqueness of each sample. A possibility could be to
estimate their "distance" from
the self-similar (unstable) fixed points of the transformation. 

We thank A. Flammini, S.
Havlin, V. Loreto, H. A. Makse and A. Vespignani for 
discussions and feedback on the manuscript. We also thank an anonymous referee for
suggesting a closer analysis of the relation between our findings and the standard renormalization
of statistical mechanics models.

\end{document}